\documentclass[sigconf]{acmart}

\AtBeginDocument{%
  }

\setcopyright{acmlicensed}
\copyrightyear{2025}
\acmYear{2025}
\acmDOI{XXXXXXX.XXXXXXX}

\acmConference[ICSE '26]{the 48th IEEE/ACM International Conference on Software Engineering}{April 12--18, 2026}{Rio de Janeiro, Brazil}
\acmISBN{978-1-4503-XXXX-X/18/06}

\usepackage{multirow}
\usepackage{tcolorbox}
\usepackage{colortbl}
\usepackage{mdframed}
\usepackage{xcolor}
\usepackage{float}
\usepackage{bbding}
\usepackage{subcaption}

\begin{document}

\title{An Empirical Study of Knowledge Distillation for Code Understanding Tasks}

\author{Ruiqi Wang}
  \authornote{Both authors contributed equally to this research.}
\affiliation{
  \institution{Harbin Institute of Technology, Shenzhen}
  \country{China}
}
\email{24s151158@stu.hit.edu.cn}

\author{Zezhou Yang}
\authornotemark[1]
\affiliation{
  \institution{Harbin Institute of Technology, Shenzhen}
  \country{China}
}
\email{yangzezhou@stu.hit.edu.cn}

\author{Cuiyun Gao}
  \authornote{Corresponding author.}
\affiliation{%
  \institution{Harbin Institute of Technology, Shenzhen}
  \country{China}
}
\email{gaocuiyun@hit.edu.cn}

\author{Xin Xia}
\affiliation{
  \institution{Zhejiang University}
  \country{China}
}
\email{xin.xia@acm.org}

\author{Qing Liao}
\affiliation{%
  \institution{Harbin Institute of Technology, Shenzhen}
  \country{China}
}
\email{liaoqing@hit.edu.cn}

\renewcommand{\shortauthors}{Ruiqi Wang, Zezhou Yang, Cuiyun Gao, Xin Xia, Qing Liao}

\begin{abstract}
Pre-trained language models (PLMs) have emerged as powerful tools for code understanding. However, deploying these PLMs in large-scale applications faces practical challenges due to their computational intensity and inference latency. Knowledge distillation (KD), a promising model compression and acceleration technique, addresses these limitations by transferring knowledge from large teacher models to compact student models, enabling efficient inference while preserving most of the teacher models' capabilities. While this technique has shown remarkable success in natural language processing and computer vision domains, its potential for code understanding tasks remains largely underexplored.

In this paper, we systematically investigate the effectiveness and usage of KD in code understanding tasks. Our study encompasses two popular types of KD methods, i.e., logit-based and feature-based KD methods, experimenting across eight student models and two teacher PLMs from different domains on three downstream tasks. The experimental results indicate that KD consistently offers notable performance boosts across student models with different sizes compared with standard fine-tuning. Notably, code-specific PLM demonstrates better effectiveness as the teacher model. Among all KD methods, the latest feature-based KD methods exhibit superior performance, enabling student models to retain up to 98\% teacher performance with merely 5\% parameters. Regarding student architecture, our experiments reveal that similarity with teacher architecture does not necessarily lead to better performance. We further discuss the efficiency and behaviors in the KD process and inference, summarize the implications of findings, and identify promising future directions.
\end{abstract}



\received{20 February 2007}
\received[revised]{12 March 2009}
\received[accepted]{5 June 2009}

\maketitle

\section{Introduction}
Automated code understanding has emerged as a critical component in software engineering, enabling machines to analyze and process source code to facilitate various downstream tasks. By transforming code into continuous vector spaces that preserve semantic and structural information, these techniques have revolutionized capabilities in clone detection \cite{DBLP:conf/wcre/WangLM0J20}, code search \cite{cocosoda,code-search-survey}, vulnerability detection \cite{DBLP:conf/icse/WenCGZZL23,DBLP:conf/issta/ChenGY0024}, and other software engineering tasks \cite{DBLP:conf/nips/LuGRHSBCDJTLZSZ21,DBLP:journals/corr/abs-2107-03374,DBLP:journals/corr/abs-2501-13742}. 

In recent years, pre-trained language models \cite{DBLP:conf/iclr/GuoRLFT0ZDSFTDC21,DBLP:conf/emnlp/0034WJH21} have become the dominant paradigm in code understanding tasks, achieving remarkable success across various downstream tasks \cite{DBLP:conf/qrs/ChenYLLG22}. These models, such as CodeBERT \cite{DBLP:conf/emnlp/FengGTDFGS0LJZ20} and UniXcoder \cite{DBLP:conf/acl/GuoLDW0022},
leverage Transformer architectures \cite{transformer} trained on large code corpora to capture syntactic and semantic relationships in programming languages, with advanced capabilities in program understanding \cite{DBLP:conf/icse/WangJLYX0L22}.
Furthermore, some general-purpose models like ModernBERT \cite{DBLP:journals/corr/abs-2412-13663} are trained for both text and code processing, also demonstrating outstanding code understanding capability. However, these models typically feature hundreds of millions of parameters, making them resource-intensive to deploy in real development environments, leading to notably slower inference.
This is particularly problematic 
since developers often require real-time responsiveness and cannot tolerate delays of seconds or minutes during basic development activities \cite{DBLP:conf/sigsoft/WangHGJ0HLD23, DBLP:conf/www/MelicherFB021}.
Consequently, there is growing interest in developing models with strong code understanding capability and minimized computational costs.

To balance the performance and efficiency of models, knowledge distillation has emerged as a particularly promising approach.
Unlike other compression and acceleration techniques such as pruning \cite{DBLP:conf/nips/MichelLN19,DBLP:conf/iclr/FrankleC19} and
quantization 
\cite{DBLP:journals/corr/abs-2303-05378,DBLP:conf/acl/TaoHZSJLLW22}, 
knowledge distillation \cite{distilBERT,DBLP:journals/corr/abs-1910-10699,pkd,tinybert,DBLP:journals/corr/abs-2110-08460,DBLP:conf/kbse/Shi0XK022,DBLP:conf/emnlp/KimR16,DBLP:journals/corr/abs-2010-13002} offers a flexible framework that transfers the knowledge from a large teacher model to a smaller, more efficient student model, enabling practical deployment without considerable performance degradation.
This technique allows the student model to learn not just from ground truth but also from the representation space of the teacher model.
Knowledge distillation methods can be broadly categorized into two types: logit-based methods, which utilize only the teacher's predicted probabilities, and feature-based methods, which leverage the teacher's intermediate feature representations \cite{DBLP:conf/aaai/XuM23}.
While knowledge distillation has demonstrated substantial success in the field of natural language processing (NLP) \cite{distilBERT,tinybert} and computer vision (CV) \cite{DBLP:journals/corr/abs-1910-10699}, it has not been comprehensively explored in code understanding. The lack of relevant research 
limits the practical deployment of large models for code understanding in resource-constrained environments, or when efficiency and responsiveness are paramount.



To mitigate this gap, we systematically investigate the effectiveness of existing knowledge distillation methods 
on code understanding models. Compared to text and images, code presents unique challenges in understanding tasks due to its distinct structural properties, semantic relationships, and domain-specific characteristics, which may influence the effectiveness of knowledge distillation approaches. Therefore, it is necessary to study several fundamental aspects in this unique domain, including achievable compression ratios, the performance impact of knowledge distillation methods, and optimal distillation settings for code-specific tasks. Guided by these objectives, our empirical study addresses the following research questions (RQs):

\begin{itemize}

\item \textbf{RQ1: What is the impact of knowledge distillation on code understanding tasks?} 

We aim to assess whether knowledge distillation can achieve model compression while retaining performance 
on code understanding tasks. Specifically, we choose defect detection, clone detection, and exception classification in our experiments. We adopt the vanilla logit-based knowledge distillation method on RoBERTa \cite{liu2019roberta} student models of five different sizes from 0.6M to 124M, with two teacher models, UniXcoder (a popular code PLM) and ModernBERT (the latest general PLM). 
We evaluate the effectiveness of knowledge distillation by comparing the performance between standard fine-tuned and distilled student models.

\item \textbf{RQ2: How do different knowledge distillation methods affect the model performance for code understanding tasks?}
Recent advances in knowledge distillation have introduced various techniques to better leverage teacher knowledge and optimize the training process.
To explore the impact of different knowledge distillation methods, we evaluate four representative methods across two widely-used model architectures: RoBERTa (Transformer-based) and BiGRU (RNN-based). By examining these methods on student models of varying sizes, we aim to verify their generalizability across architectures and model scales.

\item \textbf{RQ3: How do different student models influence the effectiveness of knowledge distillation?}
While Transformers excel in code and text processing, their high computational demands may affect their efficiency as compressed models. Building upon our investigation of RoBERTa and BiGRU in RQ2, we conduct a deeper analysis of the performance of different student model architectures (Transformer-based and RNN-based). Our experiments 
provide a more comprehensive overview of the impact of student models on knowledge distillation. 

\end{itemize}


We obtain the following findings when answering the RQs:
\begin{itemize}
    \item Knowledge distillation demonstrates consistent effectiveness across different configurations, with distilled models maintaining comparable performance to teachers while outperforming standard fine-tuned models.

    \item Feature-based distillation methods outperform logit-based approaches across code understanding tasks, with domain-specific teachers yielding better student models despite their lower standalone performance.

    \item Medium-sized student models achieve optimal performance-efficiency trade-offs, with RNN-based architectures proven more effective as student models in limited sizes. 




\end{itemize}


Our contributions are summarized as follows:
\begin{itemize}

    \item To the best of our knowledge, this is the first systematic study that extensively evaluates the effectiveness of knowledge distillation and its influencing factors in multiple code understanding tasks.

    \item In addition to performance, we measure the efficiency of training with distillation and inference for comparison with standard fine-tuning, and analyze the distilled models' behaviors which contribute to their effectiveness.

    \item We provide comprehensive insights into their effectiveness and practical applications, while summarizing our findings to guide future development and research directions.

\end{itemize}
\section{Related Work}

\subsection{Pre-trained Language Models in NLP}
The training of pre-trained language model (PLM) typically consists of two phases: pre-training and fine-tuning. During pre-training, a randomly-initialized language model learns from billions or even trillions of unannotated tokens through various objectives, such as Masked Language Modeling \cite{DBLP:conf/naacl/DevlinCLT19}, Causal Language Modeling \cite{radford2018improving}, and Denoising \cite{DBLP:journals/jmlr/RaffelSRLNMZLL20}. The model is then fine-tuned on specific downstream tasks to be a domain-specific task solver. 

\subsubsection{Encoder-only PLMs}
Understanding tasks are often performed by encoder-only PLMs. BERT \cite{DBLP:conf/naacl/DevlinCLT19} is one of the first attempts to pre-train an encoder-only Transformer, featuring 110M or 340M parameters trained on BooksCorpus and Wikipedia. ELECTRA \cite{DBLP:conf/iclr/ClarkLLM20} corrupts instead of masking random tokens during pre-training, and trains a discriminative model to predict whether each token is corrupted. DeBERTa \cite{DBLP:conf/iclr/HeLGC21} utilizes new technologies like the disentangled attention mechanism, an improved mask decoder, and virtual adversarial training to enhance the PLM's generalizability. 
ModernBERT \cite{DBLP:journals/corr/abs-2412-13663} is a recent PLM trained on 2T tokens with modern model optimization techniques to be fast and memory efficient. However, such PLMs are often trained solely on human language instead of code, which hinders their code understanding capability.

\subsubsection{PLMs with Decoders}
Many modern PLMs incorporate decoders, which compute attention unidirectionally for generation. GPT \cite{radford2018improving} is an early attempt at pre-training decoder-only Transformers. T5 \cite{DBLP:journals/jmlr/RaffelSRLNMZLL20} features an encoder-decoder architecture as a text-to-text framework. Recent large language models (LLMs) like Llama \cite{DBLP:journals/corr/abs-2302-13971} and DeepSeek \cite{DBLP:journals/corr/abs-2401-02954} support both code and text processing. Nevertheless, such models are expensive to invoke in large-scale code understanding tasks \cite{fan2024exploring}, and are out-of-scope for this paper.

\subsection{Tasks and Models in Code Understanding}
Code understanding is a subfield in software engineering, 
encompassing various tasks. Code search \cite{cocosoda,code-search-survey} retrieves relevant code snippets given the natural language query. Vulnerability detection (or defect detection) \cite{DBLP:conf/icse/WenCGZZL23} determines whether the given code contains vulnerabilities. Code question answering \cite{DBLP:conf/nips/LuGRHSBCDJTLZSZ21,huang2021cosqa} examines whether the given code can answer the natural language query.

Recent research efforts have focused on developing specialized PLMs for code understanding tasks.
CuBERT \cite{DBLP:journals/corr/abs-2001-00059} is an early encoder-only PLM for code, pre-trained from BERT-large.
CodeBERT \cite{DBLP:conf/emnlp/FengGTDFGS0LJZ20} is a 125M PLM trained on CodeSearchNet \cite{codesearchnet}, a large-scale corpus of 8M code-text pairs.
GraphCodeBERT \cite{DBLP:conf/iclr/GuoRLFT0ZDSFTDC21} leverages code structure to enhance code understanding capability.
UniXcoder \cite{DBLP:conf/acl/GuoLDW0022} supports both encoder-only, decoder-only, and encoder-decoder mode, allowing abstract syntax trees (ASTs) as inputs. CodeT5 \cite{DBLP:conf/emnlp/0034WJH21} leverages code semantics from developer identifiers, handling both code understanding and generation as sequence-to-sequence tasks. 


\subsection{Knowledge Distillation}
Knowledge distillation (KD) is a widely adopted technique in model compression and transfer learning, where a student model 
learns to mimic a teacher model's outputs or intermediate representations. 
According to the training objective \cite{DBLP:conf/aaai/XuM23}, most KD methods can be
classified as either logit-based KD or feature-based KD. 

\subsubsection{Logit-based KD} \label{logit-kd-desc}
Logits are the model's unnormalized outputs, convertible into probability distributions via the softmax function. Logit-based KD methods treat the teacher model as a black box, only using its logits for distillation. Vanilla KD 
\cite{DBLP:journals/corr/HintonVD15} replaces ground-truth probability distributions with the teacher's temperature-scaled logits. Decoupled Knowledge Distillation (DKD) \cite{DBLP:conf/cvpr/ZhaoCSQL22} splits KD loss into target class KD (TCKD) and non-target class KD (NCKD), weighting NCKD more heavily. Multi-level Logit Distillation \cite{DBLP:conf/cvpr/JinWL23} aligns teacher and student logits on sample, batch, and class levels, allowing the student model to learn higher-level representations. 

For code understanding tasks, Compressor \cite{DBLP:conf/kbse/Shi0XK022} uses a genetic algorithm to find optimal model hyperparameters before KD. Its successor Avatar \cite{DBLP:conf/icse/Shi0KXH024} prunes the hyperparameter space with an SMT solver, fine-tunes hyperparameters in a similar way to Compressor, and proceeds with KD. However, both methods focus on hyperparameter tuning and simply adopt vanilla KD.

When applied to generation tasks, logit-based KD is also known as Token-level KD, computing the average loss of all tokens. However, Sequence-level KD \cite{DBLP:conf/emnlp/KimR16} is more commonly seen in publications, where we replace ground-truth outputs with teacher-generated outputs and leave the fine-tuning procedure unchanged. 

\subsubsection{Feature-based KD}
Feature-based KD methods treat the teacher model as a white box, leveraging its intermediate representations\footnote{Although some authors merge these representations into higher-level features for distillation and refer to such methods as relation-based KD, we still categorize them as feature-based KD.} (features) for distillation. MiniLM \cite{DBLP:conf/nips/WangW0B0020,DBLP:conf/acl/WangBHDW21} trains student models with the teacher's self-attention. ReviewKD \cite{DBLP:conf/cvpr/Chen0ZJ21} aligns a student layer's representations with multiple teacher layers' representations. SimKD \cite{DBLP:conf/cvpr/ChenMZWF022} aligns the final layers of both models by reusing the teacher classifier and adding a projector.

Feature-based KD for generation tasks is rarely seen in literature, due to its complexity and the inaccessibility of intermediate representations of many commercial LLMs.

\subsubsection{Other KD methods}
BERT-of-Theseus \cite{DBLP:conf/emnlp/XuZGWZ20} progressively replaces BERT sub-modules with their compact variants. AD-KD \cite{DBLP:conf/acl/WuCQWW23} computes token-level attribution loss via integrated gradients. ReptileDistil \cite{DBLP:conf/coling/MaWYZ22} performs meta learning on the teacher model to maximize the student model's performance. Compared to previous methods, it removes the need to compute second derivatives.

While these KD methods are proven effective in the NLP and CV domains, their potential in code understanding remains underexplored. Although vanilla KD has demonstrated promise in this domain \cite{DBLP:conf/kbse/Shi0XK022,DBLP:conf/icse/Shi0KXH024}, comprehensive and systematic studies evaluating diverse KD methods for code understanding are still missing.

\section{Studied Knowledge Distillation Methods}\label{KD_description}
\begin{figure}
    \centering
    \includegraphics[width=0.5\textwidth]{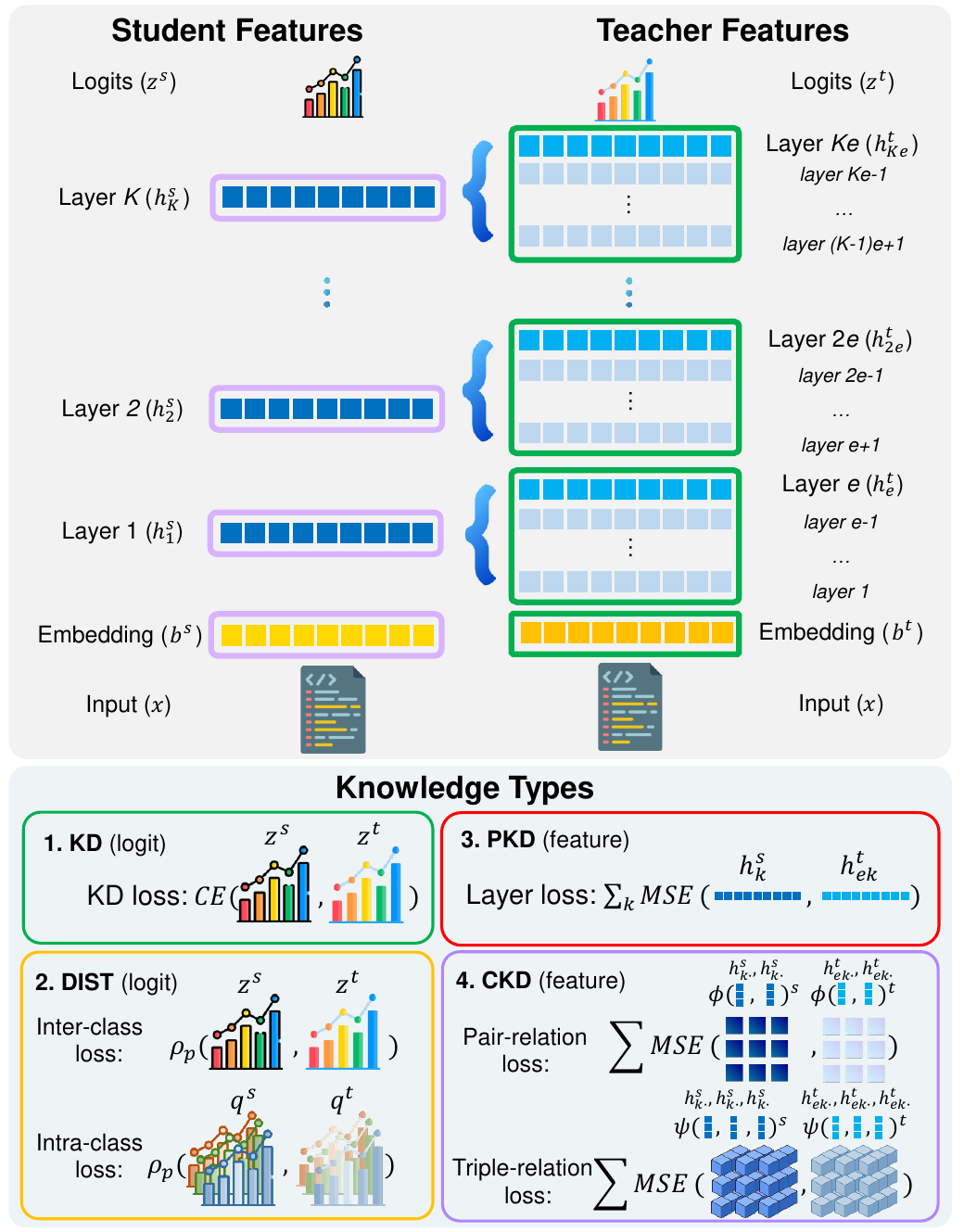}
    \caption{Overview of studied KD methods. We assume that the student and teacher models have \(K\) and \(K\times e\) layers respectively.}
    \label{fig:kd_methods}
    \vspace{-1.5em}
\end{figure}

In this section, we elaborate on four selected knowledge distillation methods in our experiments. These methods are representative in the logit-based and feature-based categories respectively. We present an overview of these methods in Figure \ref{fig:kd_methods}.

For code understanding tasks, given an input \(x\), its ground-truth distribution\footnote{A one-hot vector with the element at the ground-truth label as 1.} \(y\), and logits \(z^s\) produced by the student model, the classification loss \(L_{FT}\) for standard fine-tuning (\textbf{FT}) is computed with cross-entropy (CE):
$$
p_i=\frac{\exp z_i^s}{\sum_k\exp z_k^s},~L_{FT}=CE(p,y)=-\sum_ky_k\log p_k.
$$

For logit-based methods, we study Vanilla Knowledge Distillation ({\bf KD}) and Patience Knowledge Distillation ({\bf PKD}). For feature-based methods, we study Distillation from a Stronger Teacher ({\bf DIST}) and Contextual Knowledge Distillation ({\bf CKD}). 

\subsection{Logit-based KD}

\textbf{Vanilla Knowledge Distillation (KD)}
\cite{DBLP:journals/corr/HintonVD15} replaces the ground-truth distribution \(y\) with the softened probability distribution from the teacher model, which is obtained by first scaling the teacher's logits \(z^t\) with a temperature hyperparameter \(T\ge 1\), and then applying the softmax function:
$$
    p_i^s=\frac{\exp(z_i^s/T)}{\sum_k\exp(z_k^s/T)},~p_i^t=\frac{\exp(z_i^t/T)}{\sum_k\exp(z_k^t/T)},
$$
$$
L_{KD}=CE(p^s,p^t)=-\sum_kp^t_k\log p^s_k.
$$

\textbf{Distillation from a Stronger Teacher (DIST)} \cite{DBLP:conf/nips/0020Y00022} improves logits utilization by considering both inter- and intra-class probabilities as \(L_{DIST}=L_{inter}+L_{intra}\). The inter-class loss \(L_{inter}\) is similar to KD loss, but computes the Pearson correlation coefficient \(\rho_p\) instead of cross-entropy, allowing a relaxed match instead of identical match. The intra-class loss \(L_{intra}\) considers the logits of samples from the same class, i.e. measures the intra-class variance of semantic similarities. They are calculated as follows:
$$
L_{inter}=\rho_p(p^s,p^t),~L_{intra}=\rho_p(q^s,q^t).
$$
Here, \(q^s\) and \(q^t\) refer to the probabilities 
of intra-class samples from the student model and the teacher model, respectively.

\subsection{Feature-based KD}
Let \(h^s_{ki}\) and \(h^t_{ki}\) denote the hidden state vectors of the \(i\)-th token at the \(k\)-th layer of the student and teacher models, respectively. We assume that both models share the same hidden size. In case of different hidden sizes, we apply a simple linear transformation \(h^t_{ki}\leftarrow h^t_{ki}W_k+b_k\) before computing the loss.

Before distillation, we map each student layer \(k\) to a teacher layer \(f(k)\), enabling each student layer to learn from the representations of its corresponding teacher layer during distillation. While various mappings exist, we define \(f(k)=ek\), where the student model has \(K\) layers and the teacher model has \(e\) times as many layers, as shown in the upper half of Figure \ref{fig:kd_methods}.

{\bf Patient Knowledge Distillation (PKD)} \cite{pkd} 
directly aligns the student with the teacher through L2-normalized features using Mean-squared Error (MSE) loss:
$$
L_{PKD}=\beta\sum_{k=1}^{K-1}\|\frac{h^s_k}{\|h^s_k\|}-\frac{h^t_k}{\|h^t_{f(k)}\|}\|^2+\alpha T^2L_{KD}+(1-\alpha)L_{FT},
$$
with temperature \(T\) and weights \(\alpha\) and \(\beta\).

{\bf Contextual Knowledge Distillation (CKD)} \cite{DBLP:conf/emnlp/ParkKY21} leverages pair-wise and triple-wise token relations (TR) 
and layer-transforming relations: \(L_{CKD}=L_{KD}+\lambda(L_{TR}+L_{LTR})\), where \(\lambda\)  is the weighting coefficient for TR and LTR losses. A pair-wise relation is defined as the cosine distance while a triple-wise relation is the cosine of the angle formed by three feature points:
$$
\phi(x,y)=1-\frac{x\cdot y}{\|x\|_2\|y\|_2},~\psi(x,y,z)=\frac{(x-y)\cdot (z-y)}{\|x-y\|_2\|z-y\|_2}.
$$

TR distills the relations of tokens at different positions of the same layer. Given context length \(n\), ideally we should compute the TR relations of all \(O(n^2)\) pairs and all \(O(n^3)\) triples using both the teacher and the student models, and obtain the pair-wise loss \(L_{TR_2}\) and triple-wise loss \(L_{TR_3}\) with MSE. However, due to the computational inefficiency, we follow the original work and apply a window size \(W=10\). The overall TR loss is \(L_{TR}=L_{TR_2}+\lambda_{TR}L_{TR_3}\), with weight hyperparameter \(\lambda_{TR}\).
LTR distills the relations of tokens at the same position in different layers. The overall LTR loss \(L_{LTR}\) is computed similarly.
\section{Setup}
\label{sec:Empirical Study Setup}

In this section, we present the detailed experimental design of our study. We first fine-tune two teacher PLMs on three datasets of different code understanding tasks. Subsequently, we create eight student models, conduct knowledge distillation with two teacher PLMs on these datasets, and evaluate their performance.

\subsection{Tasks and Datasets}
To systematically evaluate the effectiveness and identify key factors of knowledge distillation, we conduct comprehensive experiments on three representative code understanding tasks.

\begin{itemize}

\item \textbf{Defect detection} requires the model to decide whether a given code snippet contains defects, such as resource leaks and null pointer dereferences. 
This challenging task necessitates sophisticated comprehension of program control and data flow dependencies.

\item \textbf{Clone detection} is a semantic similarity assessment task that determines whether two code snippets implement equivalent functionality, despite potential syntactic differences. 
This task requires
models to abstract away surface-level variations while capturing the underlying semantic behavior of both code fragments.

\item \textbf{Exception classification} involves predicting the correct exception type (such as IndexError) in Python code snippets with missing exception declarations. The task requires
models to comprehend both Python's exception-handling mechanism and potential runtime failure scenarios.

\end{itemize}

We select a widely-used dataset \textbf{Devign} \cite{DBLP:conf/nips/ZhouLSD019} for defect detection. Devign includes 27k C/C++ functions from four popular open-source projects, QEMU, FFmpeg, Wireshark, and Linux Kernel. The labels are curated by collecting security-related commits and manual labeling. We use the version provided by CodeXGLUE \cite{DBLP:conf/nips/LuGRHSBCDJTLZSZ21}, which randomly splits the dataset for training, validation, and testing with a ratio of 8:1:1.

\begin{table}[h]
    \centering
    \vspace{-0.5em}
    \caption{Model configurations. \(L\), \(H\), \(I\), \(V\), and \(C\) refer to number of layers, hidden size, intermediate size, vocabulary size, and context length respectively.}
    \begin{tabular}{c|ccccc|c}
         \toprule
         
         Model & \(L\) & \(H\) & \(I\) & \(V\) & \(C\) & Size  \\ 
         
         \midrule
         \rowcolor{gray!40}\multicolumn{7}{c}{\textbf{Teacher PLMs}}\\
         \midrule
         
         UniXcoder & 12 & 768 & 3072 & 51416 & 1026 & 126M \\
         ModernBERT & 22 & 768 & 1152 & 50368 & 8192 & 150M \\
         
         \midrule
         \rowcolor{gray!40}\multicolumn{7}{c}{\textbf{RoBERTa Student Models}}\\
         \midrule
         
         RoBERTa-124M & 12 & 768 & 3072 & 50000 & 514 & 124M \\
         RoBERTa-30M & 6 & 512 & 2048 & 20000 & 514 & 29.7M \\
         RoBERTa-7M & 4 & 384 & 768 & 5000 & 514 & 7.00M \\
         RoBERTa-1.5M & 3 & 192 & 384 & 2000 & 514 & 1.41M \\
         RoBERTa-0.6M & 3 & 128 & 256 & 1000 & 514 & 0.61M\\
         
         \midrule
         \rowcolor{gray!40}\multicolumn{7}{c}{\textbf{BiGRU Student Models}}\\
         \midrule
         
         BiGRU-30M & 6 & 448 & - & 20000 & - & 30.5M \\
         BiGRU-7M & 3 & 320 & - & 5000 & - & 7.10M \\
         BiGRU-1.5M & 2 & 176 & - & 2000 & - & 1.50M \\
         \bottomrule
         
    \end{tabular}
    \label{tab:model_config}
    \vspace{-1.5em}
\end{table}

We select the popular \textbf{BigCloneBench} \cite{DBLP:conf/icsm/SvajlenkoIKRM14} for clone detection. BigCloneBench is a large-scale dataset with over 6M samples from IJaDataset 2.0. We adopt the filtered version in CodeXGLUE with 9.1k Java snippets, forming 901k, 415k, and 415k pairs for training, validation, and testing respectively. Following their practice, we use 10\% training and validation data and all testing data.

We select \textbf{CuBERT} \cite{DBLP:journals/corr/abs-2001-00059} for exception classification. CuBERT contains several subsets of different tasks, such as function-docstring classification and variable-misuse classification. For exception classification, CuBERT provides 31k Python snippets with 20 exception types. Following CodeXGLUE's preprocessing on Devign, we split the dataset with a 8:1:1 ratio for training, validation, and testing.

We adopt a unified context length of 512 and truncate longer code, although some models allow longer context. We wrap the code with special [CLS] and [SEP] tokens, and another [SEP] token to separate the two snippets in clone detection.

\subsection{Teacher PLMs}

Our teacher PLMs are chosen to represent the state-of-the-art performance from two relevant domains: a code-specific model from software engineering, and a general-purpose model from NLP:

\begin{itemize}

\item \textbf{UniXcoder} \cite{DBLP:conf/acl/GuoLDW0022} is a unified PLM supporting multiple modes (encoder-only, decoder-only, and encoder-decoder) for code intelligence. It can process multiple input modalities (code, text, ASTs) after pre-training on both understanding and generation tasks. We enable its encoder-only mode by adding [Enc] [SEP] tokens after [CLS] in the input sequence.

\item \textbf{ModernBERT} \cite{DBLP:journals/corr/abs-2412-13663} is a family of encoder-only PLMs in NLP, incorporating modern optimization techniques like GeGLU activation and Flash Attention. 
These models are pre-trained on 2T tokens and support an extended context length of 8192 tokens, demonstrating superior performance on both NLP tasks and code retrieval. Our study utilizes the 150M \textbf{ModernBERT-base} model.

\end{itemize}

For each teacher PLM, We download the corresponding checkpoint from Hugging Face\footnote{\url{https://huggingface.co/}} and remove its MLM head if present. For each task, we append a randomly-initialized classification head and perform fine-tuning.

\subsection{Student Models}

We adopt two different student architectures for our study: the dominant {\bf Transformer}, demonstrating state-of-the-art performance at large scales; the classic {\bf RNN}, potentially remaining competitive at compact sizes. We choose a representative implementation for each architecture:

\begin{itemize}
    \item \textbf{RoBERTa} \cite{liu2019roberta} is an encoder-only Transformer implementation, with subtle differences from the original Transformer \cite{vaswani2017attention} such as replacing ReLU activations with GELU. One of the teacher PLMs, UniXcoder, also adopts this architecture. 
    We utilize the Transformers library of Hugging Face and follow common practices by using the final hidden state of the [CLS] token as the code representation.

    \item \textbf{BiGRU} \cite{cho2014learning}, specifically bidirectional gated recurrent units, is an RNN implementation with gating mechanism. 
    Although Transformers have become prevalent in large-scale models, BiGRU remains widely adopted due to its competitive performance and long-term dependency modeling capabilities. 
    We implement the BiGRU model using PyTorch, where the code representation is derived by concatenating the forward and backward final hidden states of the last token.
\end{itemize}

We select model hyperparameters based on these principles:

\begin{itemize}
    \item All model sizes approximately follow a geometric series: 124M, 30M, 7M, 1.5M, and 0.6M, with each model being roughly four times the size of the next.
    \item The encoder of each model is twice the size of the embedding layer. We choose the vocabulary sizes \(V\) accordingly, and train a Byte-pair Encoding (BPE) tokenizer for each \(V\) on the training set of each task.
\end{itemize}

All student models are listed in Table \ref{tab:model_config}. Teacher PLMs are also listed for comparison. In our experiments, student models are either fine-tuned or distilled on downstream tasks without pre-training.

\subsection{Implementation Details}

We conduct ModernBERT 
experiments on an Ubuntu server with four NVIDIA A100-40GB GPUs. All other experiments are performed on another Ubuntu server with four NVIDIA V100-32GB GPUs. We use a single GPU for each training or inference session.

In our implementation, we primarily use a batch size of 32. For CKD and ModernBERT fine-tuning, we adjust the batch size to 16 and 24 respectively due to memory constraints.
We use PyTorch's AdamW optimizer \cite{loshchilov2017decoupled} with no weight decay and the default \(\epsilon\) and \(\beta\) values along with a linear learning rate scheduler with 10
\% warmup steps. To determine the optimal configuration, we conduct grid search across various combinations of training hyperparameters. During this process, we evaluate model performance on the validation set after each epoch and preserve the checkpoint that yields the best performance.

\subsubsection{Teacher PLM fine-tuning} All teacher PLMs are fine-tuned on each dataset for 5 epochs with a learning rate of \(lr\in\{2\times 10^{-5},5\times 10^{-5}\}\), resulting in three task-specific teacher models for each PLM.

\begin{table*}
\centering
\caption{\label{overall} Results comparison between the base models and the models with knowledge distillation. The base models are fine-tuned using the original datasets. The models with knowledge distillation are fine-tuned using the datasets generated by different teacher models. UX and MB refer to UniXcoder and ModernBERT, respectively. The percentages in parentheses following the metrics denote the enhancement achieved with knowledge distillation compared to the baseline. The best results from each category are shown in bold and underlined.}
\resizebox{\linewidth}{!}{
\begin{tabular}{c|cc|cc|cc|c}

\toprule 

\multirow{2}{*}{\textbf{Model}} & \multicolumn{2}{c|}{\textbf{Defect Detection}} & \multicolumn{2}{c|}{\textbf{Clone Detection}} & \multicolumn{2}{c|}{\textbf{Exception Classification}} & \multirow{2}{*}{\textbf{Average}} \\ 

& Accuracy & F1-Score & Accuracy & F1-Score & Top-1 Acc. & Top-3 Acc. \\ 
\midrule 

UniXcoder & \textbf{\underline{65.23}} & 61.88 & 98.37 & 94.10 & 82.30 & 93.94 & 82.64 \\ 

ModernBERT & 64.20 & \textbf{\underline{62.27}} & \textbf{\underline{98.93}} & \textbf{\underline{96.10}} & \textbf{\underline{85.04}} & \textbf{\underline{94.62}} & \textbf{\underline{83.53}} \\ 
\midrule

RoBERTa-124M & 62.04 & 58.64 & \textbf{\underline{97.24}} & \textbf{\underline{89.77}} & \textbf{\underline{55.77}} & 79.40 & 73.81 \\

\multicolumn{1}{c|}{\it KD w/ UX} & \textbf{\underline{64.97}} ($4.72\% \uparrow$) & \textbf{\underline{63.54}} ($8.36\% \uparrow$) & 96.84 ($0.41\% \downarrow$) & 88.75 ($1.14\% \downarrow$) & 54.96 ($1.45\% \downarrow$) & 79.43 ($0.04\% \uparrow$) & \textbf{\underline{74.75} ($1.27\% \uparrow$)} \\

\multicolumn{1}{c|}{\it KD w/ MB} & 62.41 ($0.60\% \uparrow$) & 63.15 ($7.69\% \uparrow$) & 96.63 ($0.62\% \downarrow$) & 87.96 ($2.02\% \downarrow$) & 54.55 ($2.19\% \downarrow$) & \textbf{\underline{79.98} ($0.73\% \uparrow$)} & 74.11 ($0.41\% \uparrow$)\\
\midrule

RoBERTa-30M & 62.37 & 60.73 & \textbf{\underline{97.09}} & \textbf{\underline{89.16} } & 56.16 & 79.88 & 74.23 \\ 

\multicolumn{1}{c|}{\it KD w/ UX} & \textbf{\underline{64.13} ($2.82\% \uparrow$)} & \textbf{\underline{63.86} ($5.15\% \uparrow$)} & 96.99 ($0.10\% \downarrow$) & 89.04 ($0.13\% \downarrow$) & \textbf{\underline{58.16} ($3.56\% \uparrow$)} & \textbf{\underline{82.37} ($3.12\% \uparrow$)} & \textbf{\underline{75.76} ($2.06\% \uparrow$)} \\ 

\multicolumn{1}{c|}{\it KD w/ MB} & 64.06 ($2.71\% \uparrow$) & 62.38 ($2.72\% \uparrow$) & 96.93 ($0.16\% \downarrow$) & 88.66 ($0.56\% \downarrow$) & 55.35 ($1.44\% \downarrow$)& 80.66 ($0.98\% \uparrow$) & 74.67 ($0.59\% \uparrow$) \\
\midrule

RoBERTa-7M & 60.65 & 62.35 & 96.60 & 86.92 & 55.71 & 81.62 & 73.98 \\

\multicolumn{1}{c|}{\it KD w/ UX} & 63.25 ($4.29\% \uparrow$) & \textbf{\underline{63.86}} ($2.42\% \uparrow$) & 96.56 ($0.04\% \downarrow$) & 87.64 ($0.83\% \uparrow$)& \textbf{\underline{58.28} ($4.61\% \uparrow$)} & \textbf{\underline{82.43}} ($0.99\% \uparrow$) & \textbf{\underline{75.34}} ($1.84\% \uparrow$) \\

\multicolumn{1}{c|}{\it KD w/ MB} & \textbf{\underline{64.02} ($5.56\% \uparrow$)} & 62.12 ($0.37\% \downarrow$)& \textbf{\underline{96.69} ($0.09\% \uparrow$)} & \textbf{\underline{87.72}} ($0.92\% \uparrow$) & 55.42 ($0.52\% \downarrow$) & 81.14 ($0.59\% \downarrow$) & 74.52 ($0.73\% \uparrow$)\\
\midrule

RoBERTa-1.5M & 60.21 & 61.28 & 96.00 & 85.59 & 50.35 & 77.69 & 71.85 \\ 

\multicolumn{1}{c|}{\it KD w/ UX} & \textbf{\underline{61.71} ($2.49\% \uparrow$)} & 60.44 ($1.37\% \downarrow$)& \textbf{\underline{96.05} ($0.05\% \uparrow$)} & \textbf{\underline{85.94}} ($0.41\% \uparrow$) & 51.58 ($2.44\% \uparrow$) & \textbf{\underline{79.11}} ($1.83\% \uparrow$) & \textbf{\underline{72.47}} ($0.86\% \uparrow$) \\ 

\multicolumn{1}{c|}{\it KD w/ MB} & 60.18 ($0.05\% \downarrow$) & \textbf{\underline{61.42}} ($0.23\% \uparrow$) & 95.99 ($0.01\% \downarrow$) & 85.68 ($0.11\% \uparrow$) & \textbf{\underline{51.84}} ($2.96\% \uparrow$) & 77.66 ($0.04\% \downarrow$) & 72.13 ($0.38\% \uparrow$) \\
\midrule

RoBERTa-0.6M & 58.64 & 55.37 & 94.61 & 80.96 & 47.16 & 75.05 & 68.63 \\ 

\multicolumn{1}{c|}{\it KD w/ UX} & \textbf{\underline{59.30}} ($1.13\% \uparrow$) & 59.24 ($6.99\% \uparrow$)& 94.60 ($0.01\% \downarrow$) & 80.51 ($0.56\% \downarrow$) & \textbf{\underline{48.97} ($3.84\% \uparrow$)} & \textbf{\underline{76.14} ($1.45\% \uparrow$)} & \textbf{\underline{69.79}} ($1.69\% \uparrow$) \\ 

\multicolumn{1}{c|}{\it KD w/ MB} & 59.08 ($0.75\% \uparrow$) & \textbf{\underline{59.96}} ($8.29\% \uparrow$) & \textbf{\underline{94.87}} ($0.27\% \uparrow$) & \textbf{\underline{81.78}} ($1.01\% \uparrow$) & 47.26 ($0.21\% \uparrow$) & 75.69 ($0.85\% \uparrow$) & 69.77 ($1.66\% \uparrow$)\\
\bottomrule

\end{tabular}
}
\vspace{1.0em}
\end{table*}

\subsubsection{Student Model Distillation} For student models, we conduct two types of training approaches: (1) knowledge distillation with teacher models, and (2) standard fine-tuning without teacher models for comparison. Both approaches start from randomly initialized models, and train them for 5 epochs with a learning rate of \(lr \in\{3\times 10^{-5},1\times 10^{-4},3\times 10^{-4},1\times 10^{-3}\}\). During distillation, we set temperature to \(T\in\{1,3,10\}\).
For PKD, we set \(\alpha=1\), and select \(\beta\) from \(\{3,10\}\) for defect detection, \(\{300, 1000\}\) for clone detection, and \(\{0.3,1,3,10,30,100\}\) for exception classification. 
For CKD, we use \(\lambda,\lambda_{TR}\in\{1,3,10,30,100,300,1000\}\), with additional constraints: \(\lambda\lambda_{TR}\le 30\) for defect detection and exception classification, and \(\lambda\lambda_{TR}=1000\) for clone detection.

\subsection{Model Evaluation}
We evaluate all models on the test set of each dataset. Following previous work \cite{DBLP:conf/nips/LuGRHSBCDJTLZSZ21,DBLP:journals/corr/abs-2001-00059}, we use accuracy and F1 score for defect detection and clone detection, and Top-1 and Top-3 accuracy 
for exception classification. For better readability and interpretation, we present all metrics as a percentage score (multiplied by 100).
\section{Results}

In this section, we present the results and analysis and answer the research questions:

\begin{itemize}
    \item \textbf{RQ1:} What is the impact of knowledge distillation on code understanding tasks?

    \item \textbf{RQ2:} How do different knowledge distillation methods affect the model performance for code understanding tasks?

    \item \textbf{RQ3:} How do different student models influence the effectiveness of knowledge distillation?
\end{itemize}


\subsection{RQ1: Effectiveness of KD}

To answer this RQ, we apply vanilla KD with both teacher models to RoBERTa students in five different sizes. As shown in Table \ref{overall}, our experimental results demonstrate that KD improves model performance on average compared to standard fine-tuning. Specifically, KD with UniXcoder yields improvements ranging from 0.86\% to 2.06\% in average performance for models of various sizes. The most substantial gains are observed in defect detection, where RoBERTa-124M achieves an 8.36\% F1-score improvement through UniXcoder distillation, and RoBERTa-0.6M realizes a similar 8.29\% enhancement via ModernBERT distillation. Notably, RoBERTa-7M distilled with UniXcoder retains 91.2\% of its teacher's capabilities while using only 5.6\% of the parameters, achieving an average performance of 75.34 compared to UniXcoder's 82.64. 
Surprisingly, RoBERTa-0.6M maintains 84.5\% of UniXcoder's performance with merely 0.5\% parameters. 
These results demonstrate the effectiveness of KD for compression and for transferring code understanding capabilities from large PLMs to substantially smaller models.

\begin{tcolorbox}
[width=\linewidth-2pt,boxrule=3pt,boxsep=1pt,top=1pt, bottom=1pt, left=3pt,right=3pt, colback=gray!20,colframe=gray!25]
\textbf{Finding 1:}  
Knowledge distillation generally outperforms standard fine-tuning across all model scales, enabling student models to retain 84-91\% of teacher model capabilities while using as few as 0.5\% parameters, with particularly strong improvements in defect detection.
\end{tcolorbox}

\begin{figure}
    \centering
    \begin{subfigure}{.45\columnwidth}
        \centering
        \includegraphics[width=\linewidth]{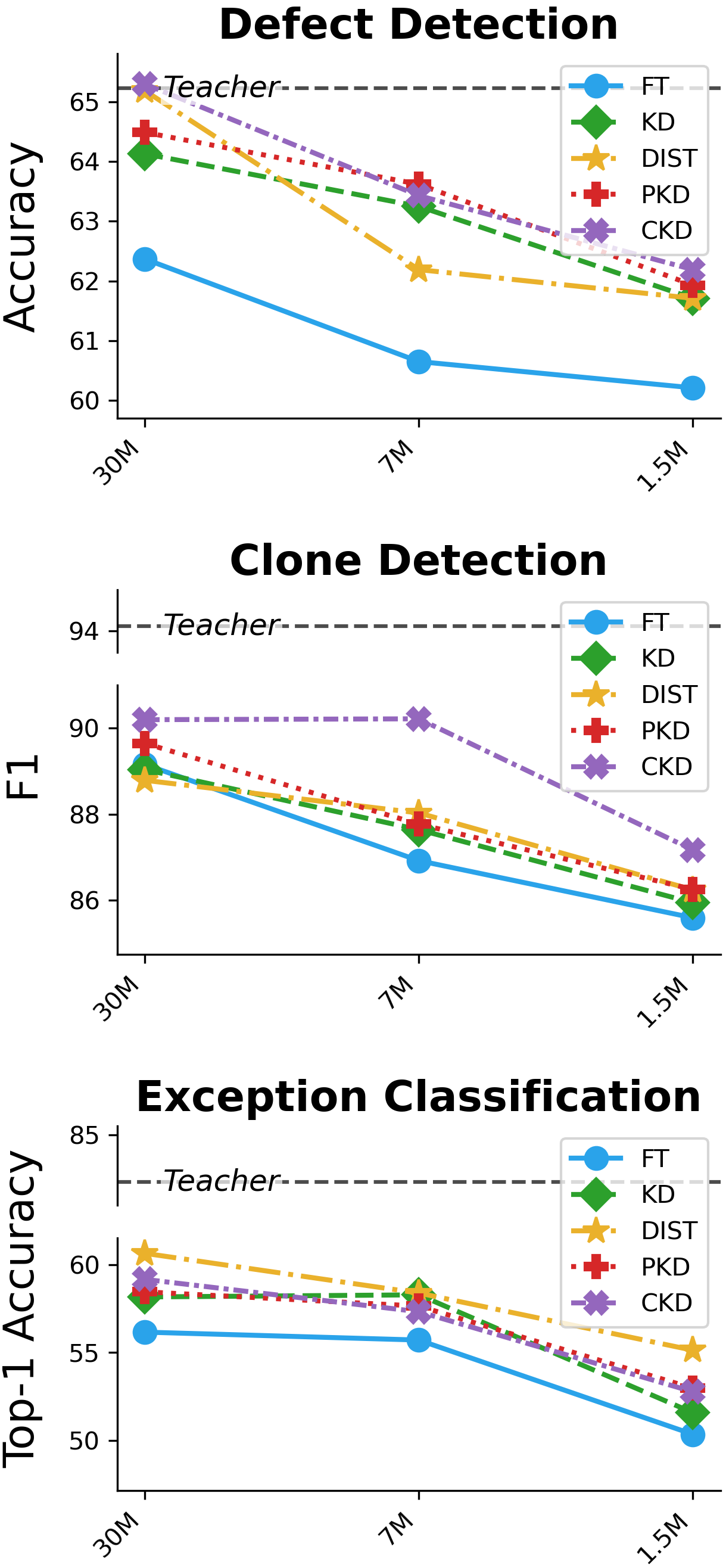}
        \caption{RoBERTa students.}
    \end{subfigure}
    \begin{subfigure}{.45\columnwidth}
        \centering
        \includegraphics[width=\linewidth]{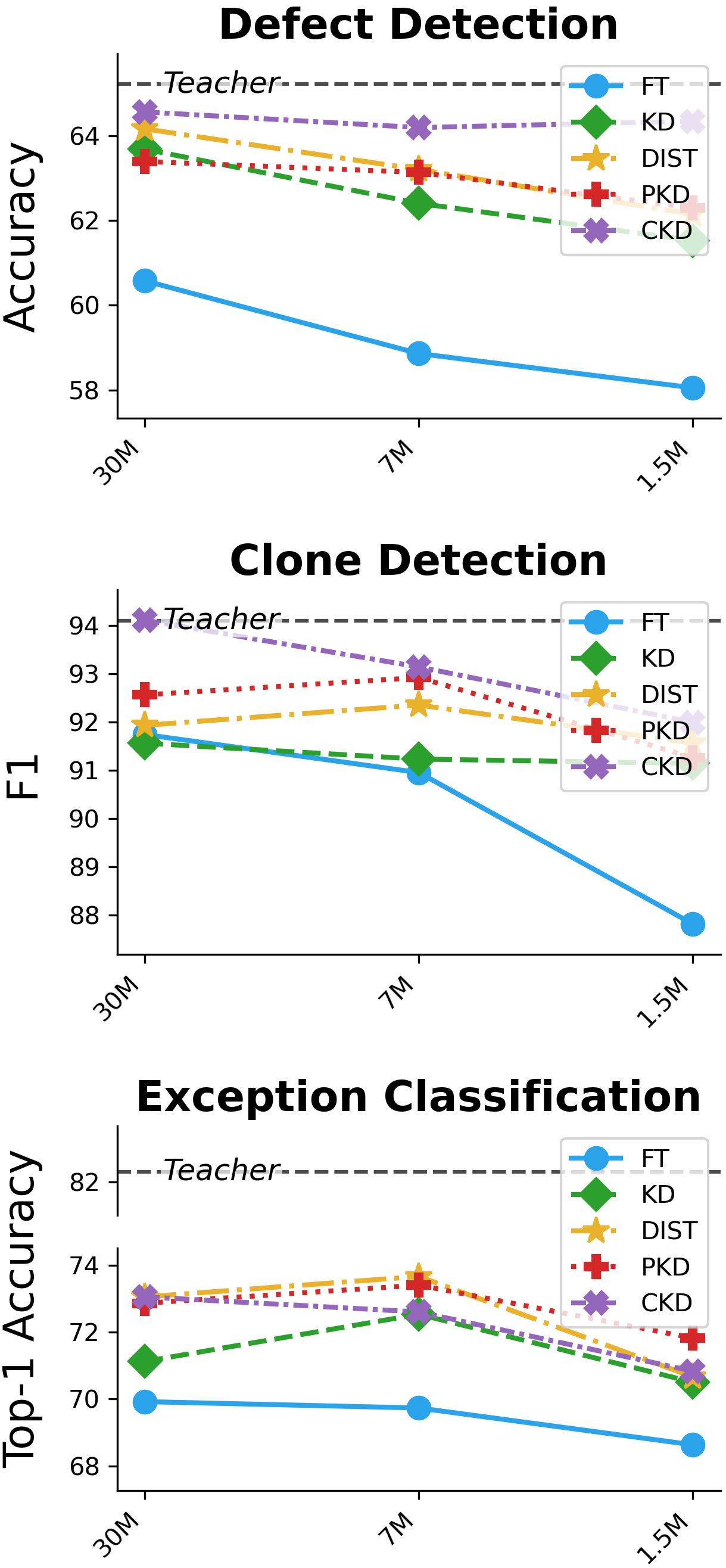}
        \caption{BiGRU students.}
    \end{subfigure}
    \caption{Performance of different student models with standard fine-tuning configuration and four KD methods for three code understanding tasks.}
    \label{fig:rq2-3}
\end{figure}

We also find a nuanced relationship between model sizes and KD effectiveness. Notably, the largest student model RoBERTa-124M exhibits modest performance enhancements from 73.81 to 74.75 (1.27\% gain) after UniXcoder distillation and to 74.11 (0.41\% gain) after ModernBERT distillation. Despite its substantial number of parameters, the 124M model underperforms relative to the more compact RoBERTa-30M when both receive knowledge transferred from UniXcoder (74.75 and 75.76 respectively). 
This trend is particularly evident in exception classification, as fine-tuned RoBERTa-124M only achieves 68\% of UniXcoder's Top-1 accuracy, while KD degrades performance by 1.45\%-2.19\%.
These results suggest that larger-scale models may require complete pre-training to acquire sufficient capabilities, as both standard fine-tuning and KD appear insufficient for their knowledge acquisition. 

The medium-scale models, such as RoBERTa-30M and RoBERTa-7M, demonstrate superior performance across different evaluation tasks. For KD with UniXcoder, the performance of RoBERTa-30M is improved by 2.06\%, from 74.23 to 75.76. Meanwhile, RoBERTa-7M achieves an average performance of 75.34 (1.84\% gain from 73.98), approaching the performance of distilled RoBERTa-30M with only 23\% parameters. 
As previously noted, distilled RoBERTa-7M maintains 91.2\% of UniXcoder's performance (82.64), suggesting a promising trade-off between compression and performance.

\begin{table}
    \caption{Average results of each student model and KD method. FT refers to fine-tuning without a teacher model.}
    \label{tab:avg_perf}

    \begin{tabular}{c|ccc|ccc}
        \toprule
        
        \multirow{2}{*}{\textbf{Method}} & \multicolumn{3}{c|}{RoBERTa} & \multicolumn{3}{c}{BiGRU} \\
        
        & 30M & 7M & 1.5M & 30M & 7M & 1.5M \\
        \midrule

        FT  & 74.23 & 73.98 & 71.85 & 77.88 & 76.78 & 75.53 \\
        \midrule
        
        KD   & 75.76 & 75.34 & 72.47 & 79.04 & 79.35 & 77.90 \\
        
        DIST & 75.81 & 75.38 & \textbf{\underline{73.94}} & 80.00 & \textbf{\underline{80.34}} & 78.43 \\
        
        PKD  & 75.90 & 75.31 & 72.87 & 79.33 & 79.54 & 78.88 \\
        
        CKD  & \textbf{\underline{76.15}} & \textbf{\underline{75.44}} & 73.21 & \textbf{\underline{80.69}} & 79.86 & \textbf{\underline{79.09}} \\
        \bottomrule
        
    \end{tabular}
\end{table}

The small-scale models demonstrate consistent albeit modest improvements through KD. Specifically, the performance of RoBERTa-1.5M increases from 71.85 to 72.47 after KD with UniXcoder, while RoBERTa-0.6M improves from 68.63 to 69.79. Notably, despite its extremely compact size, the RoBERTa-0.6M model retains approximately 84\% teacher capabilities with less than 0.5\% parameters, demonstrating the effectiveness of KD for resource-constrained scenarios. Our task-specific analysis reveals that defect detection consistently benefits most from distillation across most model scales, demonstrating accuracy improvements of 2.5-4.8\% with UniXcoder and a student model larger than 0.6M, while clone detection shows more modest gains, particularly in smaller architectures.

\begin{tcolorbox}
[width=\linewidth-2pt,boxrule=3pt,boxsep=1pt,top=1pt, bottom=1pt, left=3pt,right=3pt, colback=gray!20,colframe=gray!25]
\textbf{Finding 2:}  
Knowledge distillation exhibits a non-linear effectiveness curve across model sizes, with mid-sized architectures (7M-30M parameters) achieving optimal efficiency. 
The larger models and smaller models achieve consistent but modest improvements.
\end{tcolorbox}

Throughout our experiments, UniXcoder demonstrates superior knowledge transfer efficacy compared to ModernBERT, despite the latter's slightly higher standalone performance (83.53 and 82.64, respectively). This pattern is consistent across all student model scales and is particularly evident in exception classification, where RoBERTA-30M and 7M receive 3.6-4.6\% performance boosts when distilled with UniXcoder but 0.5-1.4\% degrade with ModernBERT. 
These results suggest that a teacher model's effectiveness in KD is not only determined by its absolute performance, but also by its knowledge transfer ability, where UniXcoder's code-specialized pre-training offers clear advantages.

\begin{tcolorbox}
[width=\linewidth-2pt,boxrule=3pt,boxsep=1pt,top=1pt, bottom=1pt, left=3pt,right=3pt, colback=gray!20,colframe=gray!25]
\textbf{Finding 3:}  
Domain-specialized teacher models outperform general-purpose architectures in knowledge transfer effectiveness 
despite lower standalone performance, enabling compressed models to retain up to 91\% of teacher capabilities while using as little as 5.6\% of the parameters.
\end{tcolorbox}

\subsection{RQ2: Impact of KD Methods}

\begin{table*}[htbp]
    \centering
    \caption{Time cost of training and inference of different KD methods across various student models,
    with UniXcoder as the teacher model. The values represent the time per step in milliseconds with the speed factor shown in parentheses.
    FT, KD, DIST, PKD, and CKD refer to standard fine-tuning, vanilla KD, distillation from a stronger teacher, patience knowledge distillation, and contextual knowledge distillation respectively.
    }
    
    \begin{tabular}{c|ccccc|c}
        \toprule
        \multirow{2}{*}{\bf Model} & \multicolumn{5}{c|}{\bf Training} & \multirow{2}{*}{\bf Inference} \\
        & FT & KD & DIST & PKD & CKD  \\
        \midrule
        UniXcoder & 848 (1.0×) & - & - & - & - & 252 (1.0×) \\
        \midrule
        RoBERTa-30M & 226 (3.8×) & 464 (1.8×) & 526 (1.6×) & 528 (1.6×) & 1460 (0.58×) & 72.5 (3.5×) \\
        RoBERTa-7M & 103 (8.2×) & 341 (2.5×) & 404 (2.1×) & 404 (2.1×) & 927 (0.91×) & 32.5 (7.8×) \\
        RoBERTa-1.5M & 52.4 (16×) & 290 (2.9×) & 353 (2.4×) & 353 (2.4×) & 636 (1.3×) & 19.4 (13×) \\
        \midrule
        BiGRU-30M & 415 (2.0×) & 723 (1.2×) & 722 (1.2×) & 725 (1.2×) & 2009 (0.42×) & 183 (1.4×) \\
        BiGRU-7M & 140 (6.1×) & 450 (1.9×) & 450 (1.9×) & 453 (1.9×) & 933 (0.91×) & 70.0 (3.6×) \\
        BiGRU-1.5M & 93.8 (9.0×) & 401 (2.1×) & 402 (2.1×) & 405 (2.1×) & 611 (1.4×) & 56.5 (4.5×) \\
        \bottomrule
    \end{tabular}
    \label{tab:efficiency}
\end{table*}
To answer RQ2 and RQ3, we apply all four KD methods with UniXcoder as the teacher model.
As shown in Figure \ref{fig:rq2-3}, the performance comparison of different tasks reveals distinct patterns in the effectiveness of KD methods. For defect detection, contextual knowledge distillation (CKD) demonstrates particularly strong performance across both architectures. For example, BiGRU-1.5M+CKD configuration achieves a remarkable 64.35 accuracy, which substantially outperforms its standard fine-tuning (FT) configuration and approaches the performance of BiGRU-30M+CKD configuration. Clone detection allows CKD to exhibit the strongest cross-architecture consistency, with F1 scores of 90.20 and 90.22 for RoBERTa-30M and RoBERTa-7M, respectively. It demonstrates that effective knowledge transfer maintains student performance even with substantial parameter reduction. For exception classification, Distillation from a Strong Teacher (DIST) appears particularly effective for the RoBERTa architecture, improving performance by 4.48 and 4.78 points over its FT configuration with 30M and 1.5M parameters.


\begin{tcolorbox}
[width=\linewidth-2pt,boxrule=3pt,boxsep=1pt,top=1pt, bottom=1pt, left=3pt,right=3pt, colback=gray!20,colframe=gray!25]
\textbf{Finding 4:}  
Different code understanding tasks benefit from specific distillation method: CKD shows remarkable stability in clone detection and defect detection across model sizes, while DIST demonstrates superior results in exception classification, particularly with RoBERTa architectures.
\end{tcolorbox}

The effectiveness of different KD methods remains consistent across architectures, with CKD and DIST generally outperforming Patient Knowledge Distillation (PKD) and vanilla KD. PKD offers better parameter-scaling properties than vanilla KD, particularly in BiGRU architectures where performance degradation from 30M to 1.5M parameters is minimized. Notably, all advanced KD methods (CKD, DIST, PKD) outperform vanilla KD, demonstrating the importance of incorporating contextual or structured objectives in the KD process when targeting ultra-compact models.

Our comparative analysis of KD methods reveals distinct effectiveness patterns across model architectures and sizes. As illustrated in Table \ref{tab:avg_perf}, all KD methods consistently outperform FT across both RoBERTa and BiGRU architectures. For RoBERTa models, CKD demonstrates superior performance on the 30M and 7M parameter configurations, achieving average scores of 76.15 and 75.44 respectively, while DIST excels with 1.5M models (73.94). The BiGRU models exhibit similar patterns, with CKD yielding the highest average performance for both 30M and 1.5M models,
while DIST slightly edges out CKD for 7M models.

The performance retention characteristics of different distillation methods reveal compelling efficiency patterns. While all models experience some performance degradation with parameter reduction, CKD demonstrates remarkable resilience, particularly for clone detection tasks where RoBERTa-1.5M maintains 96.6\% of the performance achieved by RoBERTa-30M. BiGRU architectures exhibit even stronger retention characteristics, with BiGRU-1.5M+CKD preserving approximately 98\% of BiGRU-30M. This experimental results suggests that CKD facilitate more effective representation alignment between teacher and student models, enabling smaller architectures to capture essential aspects.

\begin{tcolorbox}
[width=\linewidth-2pt,boxrule=3pt,boxsep=1pt,top=1pt, bottom=1pt,left=3pt,right=3pt, colback=gray!20,colframe=gray!25]
\textbf{Finding 5:}  
Advanced knowledge distillation methods, especically the latest feature-based KD method, demonstrate superior effectiveness over vanilla KD method across diverse model configurations.
\end{tcolorbox}

\subsection{RQ3: Impact of Student Architecture}

The BiGRU architecture demonstrates superior knowledge retention capabilities compared to RoBERTa across all distillation methods and model sizes. This is particularly evident in the exception classification task, where BiGRU-1.5M+CKD achieves 70.83 accuracy, which is substantially higher than RoBERTa-1.5M+CKD (52.77). This architecture-specific advantage persists across all three tasks, suggesting that BiGRU's recurrent inductive bias may enable more effective knowledge transfer from teacher models. The performance gap between architectures widens as model size decreases, highlighting the importance of architecture selection when targeting extreme compression scenarios.

The relationship between model size and distillation method effectiveness reveals non-linear patterns worth noting. While performance generally decreases with model size reduction, the rate of degradation varies between distillation methods. CKD and DIST demonstrate remarkable resilience to parameter reduction, with performance curves showing substantially gentler slopes than vanilla KD and FT. This trend is especially evident in clone detection, where CKD achieves nearly constant performance across model sizes for both architectures. This suggests that certain code understanding capabilities can be effectively preserved even in extremely parameter-constrained scenarios.

\begin{tcolorbox}
[width=\linewidth-2pt,boxrule=3pt,boxsep=1pt,top=1pt, bottom=1pt, left=3pt,right=3pt, colback=gray!20,colframe=gray!25]
\textbf{Finding 6:}  
The effectiveness of KD shows little correlation with architectural similarity between teacher and student models. Different student architectures exhibit similar size-performance trade-offs, suggesting that architectural compatibility is not a decisive factor in knowledge transfer.
\end{tcolorbox}

\section{Discussion}

\subsection{Training and Inference Efficiency}
We study the time costs of training and inference of each model with each KD method in Table \ref{tab:efficiency}, excluding data preparation phase with JSON parsing and tokenization as well as model loading. To ensure that all methods require the same number of training steps, the batch size of CKD is normalized from 16 to 32.

\begin{table}[]
    \centering
    \caption{Behavioral analysis of student models trained with different methods. \textbf{Agr.} and \textbf{Acc.} refers to Agreement and Accuracy respectively, while \textbf{(D)} denotes student performance on disagreeing samples.}
    \begin{tabular}{c|cc|cc|cc}
         \toprule
         \multirow{2}{*}{\bf Method} & \multicolumn{2}{c|}{\bf Defect} & \multicolumn{2}{c|}{\bf Clone} & \multicolumn{2}{c}{\bf Exception}  \\
         & Agr. & Acc. (D) & Agr. & F1 (D) & Agr. & Top-1 (D) \\
         \midrule
         \rowcolor{gray!40}\multicolumn{7}{c}{\bf RoBERTa-30M} \\
         \midrule

         FT & 73.35 & 44.64 & 97.40 & 12.56 & 57.06 & 9.23 \\
         KD & 75.40 & 47.77 & 97.12 & 14.05 & 59.22 & 9.64 \\
         CKD & 77.89 & 50.17 & 97.80 & 18.91 & 60.28 & 10.23 \\
         \midrule

         \rowcolor{gray!40}\multicolumn{7}{c}{\bf RoBERTa-7M} \\
         \midrule

         FT & 69.73 & 42.44 & 96.98 & 7.40 & 57.51 & 8.27 \\
         KD & 73.57 & 46.26 & 97.11 & 10.32 & 60.03 & 8.87 \\
         CKD & 75.59 & 46.33 & 97.85 & 17.59 & 59.41 & 8.18 \\
         \midrule

         \rowcolor{gray!40}\multicolumn{7}{c}{\bf RoBERTa-1.5M} \\
         \midrule

         FT & 68.78 & 41.97 & 96.32 & 10.50 & 51.71 & 6.68 \\
         KD & 71.74 & 43.78 & 96.54 & 10.14 & 53.74 & 6.20 \\
         CKD & 70.17 & 44.91 & 96.98 & 11.64 & 54.71 & 7.54 \\
         \midrule

         \rowcolor{gray!40}\multicolumn{7}{c}{\bf BiGRU-30M} \\
         \midrule

         FT & 70.10 & 42.23 & 98.37 & 20.46 & 71.47 & 15.14 \\
         KD & 76.28 & 46.76 & 98.35 & 21.76 & 74.34 & 14.95 \\
         CKD & 77.60 & 48.53 & 98.87 & 38.26 & 75.69 & 17.11 \\
         \midrule

         \rowcolor{gray!40}\multicolumn{7}{c}{\bf BiGRU-7M} \\
         \midrule

         FT & 68.67 & 39.84 & 98.00 & 16.99 & 72.40 & 13.32 \\
         KD & 73.10 & 44.76 & 98.16 & 17.80 & 76.21 & 14.36 \\
         CKD & 76.57 & 47.81 & 98.70 & 27.43 & 75.66 & 15.50 \\
         \midrule

         \rowcolor{gray!40}\multicolumn{7}{c}{\bf BiGRU-1.5M} \\
         \midrule

         FT & 66.98 & 39.14 & 96.99 & 11.83 & 71.15 & 13.63 \\
         KD & 73.02 & 43.15 & 98.19 & 9.55 & 74.21 & 13.25 \\
         CKD & 75.92 & 48.18 & 98.38 & 20.68 & 74.40 & 13.60 \\
         \bottomrule

    \end{tabular}
    \label{tab:behavior}
    \vspace{-2.0em}
\end{table}

{\bf KD slows down training remarkably, though most KD methods train the student model at comparable speeds.} For compressed models ranging from 1.5M to 30M parameters, standard fine-tuning offers a \(2\times\) to \(16\times\) speedup, contributed solely by forward and backward passes. In contrast, KD, PKD, and DIST at least double the training time, as the teacher model's inference accounts for a large portion, especially for smaller student models. CKD requires substantially longer training time for two reasons: (1) smaller batch sizes leading to more training steps, and (2) more computationally expensive loss function and backward passes.

{\bf As models become smaller, speed increases proportionally until reaching tiny sizes, after which diminishing returns are observed.} For instance, compressing UniXcoder (126M) into RoBERTa-30M (or 7M) yields approximately \(3.7\times\) (or \(8.0\times\)) speedups, but further compression from 7M to 1.5M results in a speedup of less than \(2\times\). A similar trend is observed for BiGRU models, although their RoBERTa counterparts are substantially faster due to the highly parallelizable Transformer architecture and optimizations of modern GPUs.

\subsection{Behavioral Impact of KD}
To investigate how knowledge distillation improves model performance, we analyze the behavioral differences between distilled and standard fine-tuned student models. The experimental results are shown in Table \ref{tab:behavior}.
Specifically, we utilize two metrics to analyze the alignment between teacher and student models: (1) \textbf{Agreement}, which measures the proportion of samples where the student and teacher models make the same prediction, and (2) \textbf{Performance on Disagreement}, which evaluates the student model's effectiveness on samples where its predictions differ from the teacher using standard metrics like accuracy and F1 scores. We make the following discoveries:

\textbf{KD increases Agreement between student and teacher models compared to standard fine-tuning}. Vanilla KD offers an Agreement boost ranging from 2.8\% for RoBERTa-30M to 9.0\% for BiGRU-1.5M on defect detection, and a more consistent 3.8\%-5.3\% improvement on exception classification. CKD provides an even larger Agreement enhancement of up to 13.3\% for BiGRU-1.5M on defect detection, demonstrating its superior ability to align student models with the teacher. Nevertheless, there is still room for further improvement, likely due to the limited capacities of student models and the absence of pre-training. 

\textbf{KD improves the student model's performance even on samples where it disagrees with the teacher.} For defect detection task, vanilla KD enhances accuracy on disagreeing samples by 7.0\%-23.1\%. For clone detection task, despite the already saturated Agreement, vanilla KD achieves a 138\% (from 7.40 to 17.59) improvement for RoBERTa-7M on disagreeing samples. In contrast, the improvement on exception classification is modest, likely due to the task's inherent difficulty for student models, as evidenced by lower Agreement and extremely low accuracy on disagreeing samples. This suggests a need for more effective compressed architectures and training methods. Finally, CKD outperforms vanilla KD with greater enhancements as anticipated.

In summary, KD enhances student model performance through two primary mechanisms: increasing prediction alignment with teacher models and improving accuracy on samples where student-teacher predictions differ. The effectiveness of these mechanisms varies across tasks and model architectures, with more complex task (e.g., exception classification) showing limited improvements particularly on disagreeing samples.

\subsection{Quantitative Analysis of Student-Teacher Alignment}
To further explore how knowledge distillation affects the alignment between student and teacher models, we conduct a quantitative analysis using two metrics:
ground-truth (GT) negative log-likelihood (NLL) error and non-GT entropy error. 
The first metric quantifies the confidence gap between the student and teacher models on the correct answer, while the second measures the disparity in uncertainty across incorrect answers \cite{DBLP:conf/cvpr/ZhaoCSQL22}.
Let \(p^s\) and \(p^t\) denote the predicted distributions from the student and teacher models, respectively. We define the metrics as follows:

\begin{enumerate}
    \item {\bf GT NLL error.} 
    Given the predicted probability \(p_i\) of class \(i\), its NLL is \(-\log p_i\). The GT NLL error between the student and teacher models can be computed as follows:
    $$
        \Delta(\text{NLL})=|\text{NLL}^s-\text{NLL}^t|=|\log p^s_C-\log p^t_C|,
    $$
    where \(C\) is the GT class.
    
    \item {\bf Non-GT entropy error.} 
    Given the predicted probability \(p_i\) of class \(i\) and the GT class \(C\), we first normalize the non-GT distributions:
    $$
    q_i^s=\frac{p^s_i}{1-p_C^s},q_i^t=\frac{p^t_i}{1-p_C^t}, \forall i\ne C.
    $$
    We then compute their entropy values \(H(q^s), H(q^t)\) for these two normalized distributions. For a given distribution \(q\), its entropy is calculated as:
    $$
    H(q)=-\sum_{i\ne C}q_i\log q_i.
    $$
    Finally, the student-teacher alignment in non-GT classes can be measured by the absolute difference between their entropy values:
    $$
    \Delta(\text{Entropy})=|H(q^s)-H(q^t)|.
    $$
\end{enumerate}

\begin{figure}
    \centering
    \begin{subfigure}{.45\columnwidth}
        \centering
        \includegraphics[width=\linewidth]{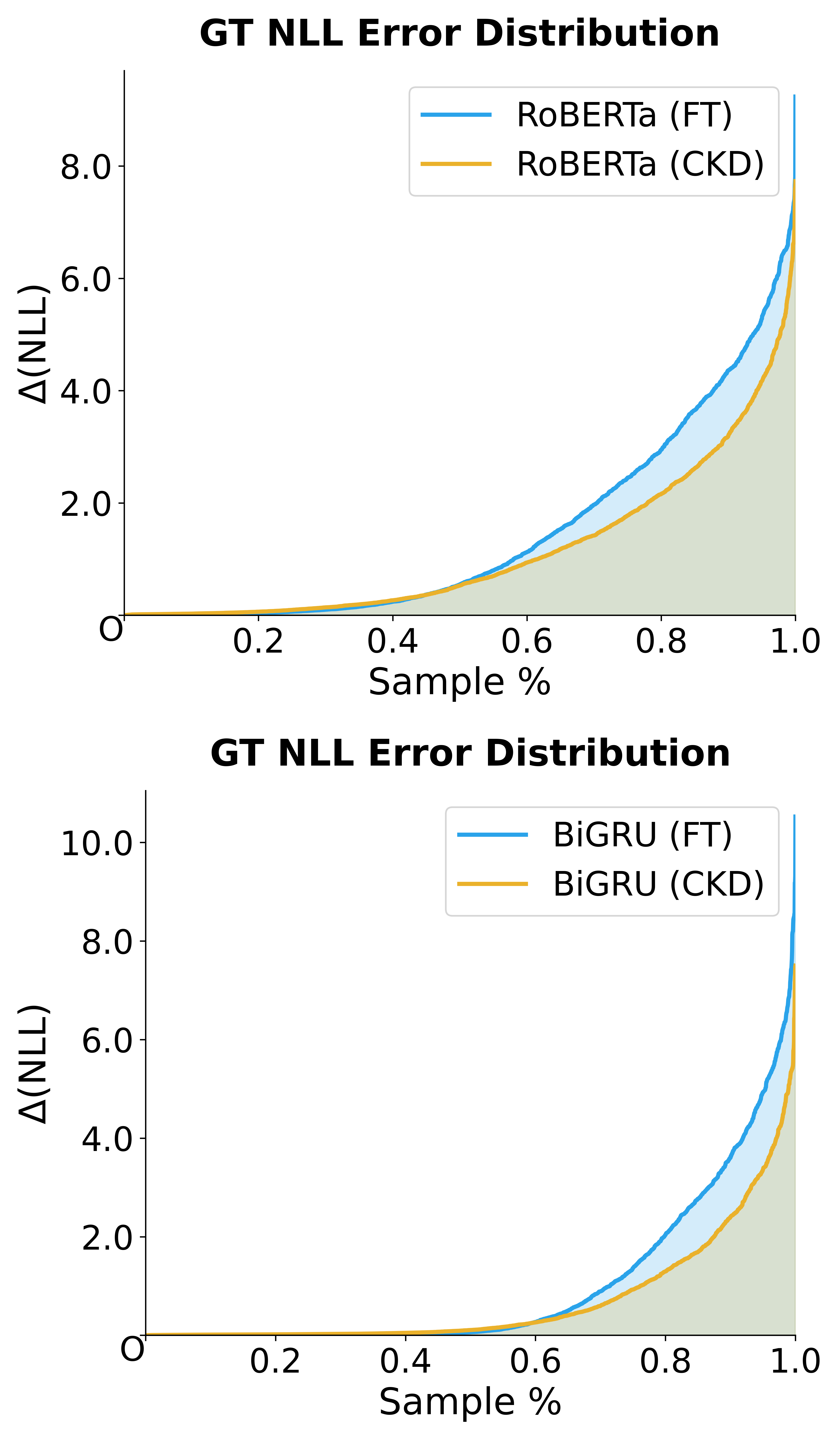}
        \caption{GT NLL errors.}
    \end{subfigure}
    \begin{subfigure}{.45\columnwidth}
        \centering
        \includegraphics[width=\linewidth]{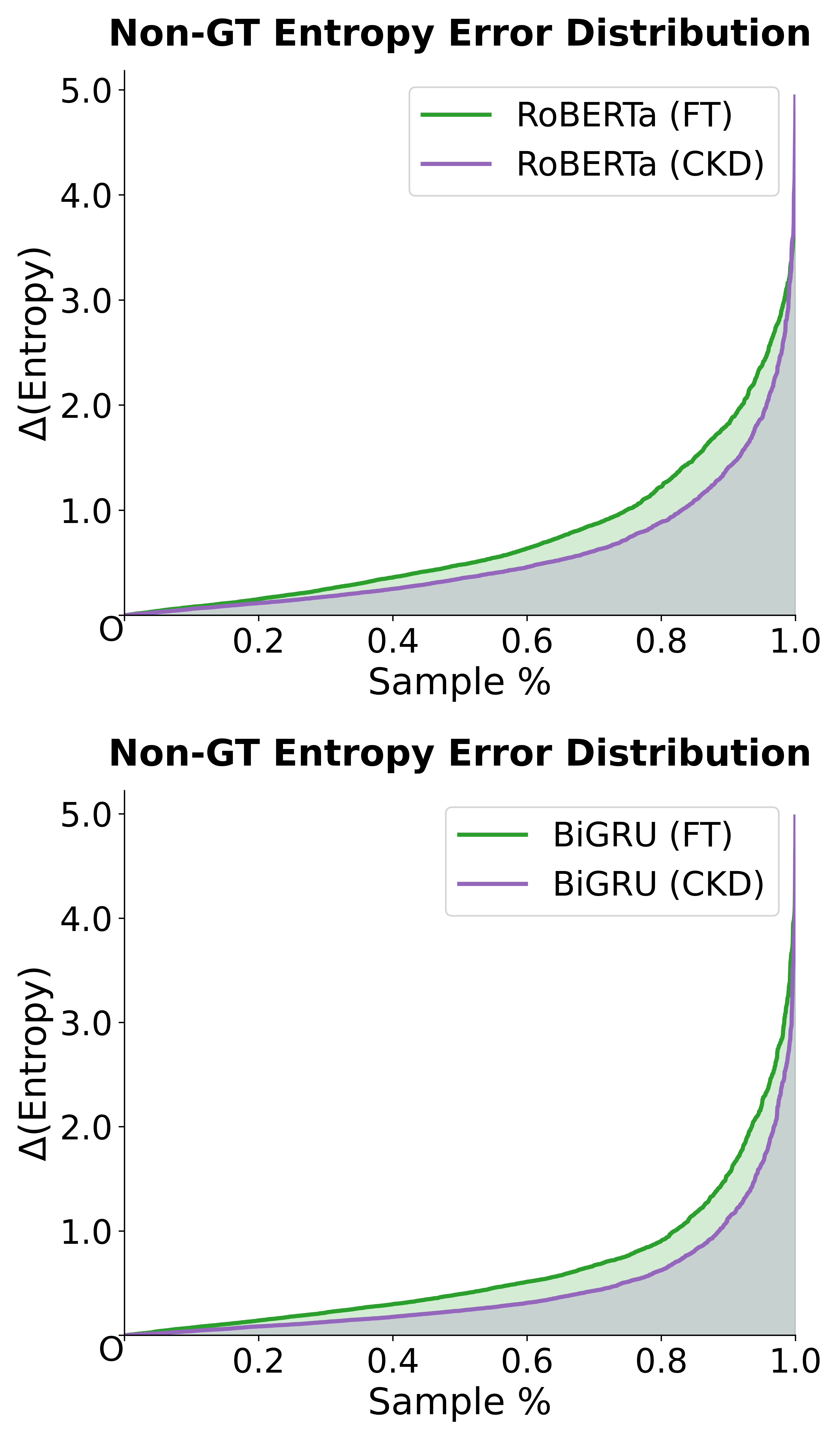}
        \caption{Non-GT entropy errors.}
    \end{subfigure}
    \caption{The impact of FT and CKD on student-teacher alignment across different sample percentiles. GT refers to the ground-truth class.}
    \label{fig:align}
\end{figure}

We choose RoBERTa-30M and BiGRU-30M as the representative student models, and render the distributions of the above metrics on exception classification with kernel density estimation in Figure \ref{fig:align}. We compare standard fine-tuned student models with those trained via CKD using UniXcoder as the teacher. 

As shown in Figure \ref{fig:align}, student models distilled with CKD consistently exhibit lower errors than their fine-tuned counterparts across the entire sample range for both GT NLL and non-GT entropy errors. The difference in performance becomes more pronounced as the sample percentile increases, particularly in the upper percentiles where the errors are largest. While fine-tuned students show higher error rates in these upper percentiles, the distilled students maintain a closer alignment to the teacher model. This improved alignment can be observed in both the GT class predictions, which relate to the correct answers, and the non-GT class predictions, which provide information about incorrect options. These findings suggest that knowledge distillation enhances the student model's ability to mimic the teacher's predictions across various types of samples, contributing to the overall performance improvement observed after distillation.

\subsection{Implications of Study}
We summarize the implications of our empirical study from both  developers' and researchers' perspectives as follows.

{\bf For developers:} Our findings indicate that KD yields better student models than standard fine-tuning, allowing model compression and inference acceleration while retaining up to 90\% teacher performance for code understanding, at the cost of longer training time. We further conclude the following insights:

\begin{enumerate}
    \item Developers should carefully select distillation methods based on expected performance and training budget.

    \begin{itemize}
        \item Vanilla KD serves as a simple yet effective choice to improve student model performance for practical deployment scenarios requiring model compression and acceleration. 

        \item While advanced KD methods can achieve better performance, they often require more computational resources. Developers should carefully weigh these performance gains against increased resource demands.
        
    \end{itemize}

    \item Domain-specific models like UniXcoder serve as better teachers in the process of  KD compared to state-of-the-art general-purpose models like ModernBERT for code understanding tasks.

    \item The choice of student model architecture requires careful consideration. Our findings suggest that architectural similarity between teacher and student models does not guarantee better KD performance. Instead, student models with strong inherent capabilities may yield greater benefits.

    
\end{enumerate}

{\bf For researchers:} Our study reveals the effectiveness and limitations of current KD methods in code understanding tasks, suggesting potential future directions:

\begin{enumerate}
    \item Most KD methods are applied exclusively during fine-tuning phase, omitting pre-training phase entirely, resulting in suboptimal model performance, particularly for larger and more capable student models like RoBERTa-30M and 124M. To address this, researchers could explore additional distillation stages such as the KD methods within pre-training stage with a teacher model, to avoid fine-tuning from scratch and achieve better results.

    \item Current KD methods are tailored for text and image processing instead of code understanding. To bridge this gap, researchers could explore more robust approaches that incorporate code structure information such as abstract syntax trees (ASTs) and control-flow graphs (CFGs) for better code understanding. Furthermore, it is crucial to develop novel KD methods applicable to a broader range of software engineering tasks.

    \item While LLMs have been widely adopted for code intelligence tasks, their unified generation approach for various tasks makes it challenging to directly apply existing knowledge distillation methods. Therefore, developing efficient, compact code-specific models through KD remains an urgent challenge.

    \item Previous studies have primarily focused on the effectiveness of KD itself for model compression, acceleration, and performance retention. Future work could explore fusing KD with other compression approaches like pruning and quantization to prioritize efficiency, or synthesizing multiple student models for inputs of varying difficulties to maximize performance.
    
\end{enumerate}

\subsection{Threats to Validity}
We identify the following threats to validity of our study:

\begin{itemize}
    \item {\bf Limited datasets.} Our experiments are conducted on three code understanding task datasets. However, the limited number of datasets may introduce bias to the results and limit the generalizability of our findings. We mitigate the bias by adopting a popular and representative dataset for each task, and enhance generalizability by experimenting on multiple student models and with four KD methods.

    \item {\bf Result reproducibility.} Our experimental results could be influenced by differences in hardware and model hyperparameters, which pose challenges to reproducibility. To address this, we make our code for distillation and inference publicly available\footnote{\url{https://github.com/BackOnTruck/kd-empirical}}, and disclose the technical details including hyperparameters in our paper. This ensures transparency and facilitates replication studies by other researchers, enhancing the reliability of our findings.


    \item {\bf Distillation of LLMs.} 
    We investigate the effectiveness of knowledge distillation on two PLMs instead of LLMs, which may limit the comprehensiveness of our findings.
    However, directly incorporating LLMs into our empirical framework would be inappropriate, as LLMs process code understanding tasks by generating tokens instead of producing logits, and they involve numerous features that complicate the distillation process. 
    In the future, we will explore specialized KD methods designed for LLMs in code intelligence applications.
\end{itemize}
\section{Conclusion}
In this paper, we empirically explore the effectiveness and applications of four knowledge distillation (KD) methods on three code understanding tasks. 
Our experiments are conducted across two teacher models and eight student architectures, demonstrating that KD consistently outperforms standard fine-tuning across different model configurations. Our findings also reveal that code-specific PLMs serve as more effective teachers, and the latest feature-based KD methods can achieve superior performance. Notably, the similarity between student and teacher architectures does not emerge as a decisive factor for KD effectiveness. 
We further discuss the efficiency of KD training and inference, analyze student behavior changes, and summarize the implications, expecting future development and research to benefit from our insights.


\bibliographystyle{ACM-Reference-Format}
\bibliography{references}

\end{document}